  \documentclass[aps]{revtex4-1}
\usepackage{graphicx,amsmath}\usepackage{color}
\draft % marks overfull lines with a black rule on the right

\begin{document}

% Use the \preprint command to place your local institutional report number
% on the title page in preprint mode.
% Multiple \preprint commands are allowed.
%\preprint{}

\title{Barut-Girardello coherent states for nonlinear oscillator with position-dependent mass } %Title of paper

% repeat the \author .. \affiliation  etc. as needed
% \email, \thanks, \homepage, \altaffiliation all apply to the current author.
% Explanatory text should go in the []'s,
% actual e-mail address or url should go in the {}'s for \email and \homepage.
% Please use the appropriate macro for the type of information

% \affiliation command applies to all authors since the last \affiliation command.
% The \affiliation command should follow the other information.

 \author{Naila Amir}
 \email[]{naila.amir@live.com, naila.amir@seecs.edu.pk}
% %\homepage[]{Your web page}
% %\thanks{}
% %\altaffiliation{}
 \affiliation{School of Electrical Engineering and Computer Sciences,
 	National University of Sciences and Technology,
 	H-12, Islamabad, Pakistan.}
\author{Shahid Iqbal}
\email[]{sic80@hotmail.com, siqbal@sns.nust.edu.pk}
%\homepage[]{Your web page}
%\thanks{}
%\altaffiliation{}
\affiliation{School of Natural Sciences, National University of Sciences and
             Technology, H-12, Islamabad, Pakistan}

% Collaboration name, if desired (requires use of superscriptaddress option in \documentclass).
% \noaffiliation is required (may also be used with the \author command).
%\collaboration{}
%\noaffiliation

\date{\today}

\begin{abstract}
Using ladder operators for the non-linear oscillator with position-dependent effective mass, realization of the dynamic group $SU(1,1)$ is presented. Keeping in view the algebraic structure of the non-linear oscillator, coherent states are constructed using Barut-Girardello formalism and their basic properties are discussed. Furthermore, the statistical properties of these states are investigated by means of Mandel parameter and second order correlation function. Moreover, it is shown that in the harmonic limit, all the results obtained for the non-linear oscillator with spatially varying mass reduce to corresponding results of the linear oscillator with constant mass.

\end{abstract}

 % insert suggested PACS numbers in braces on next line
% 03.65.Ta  Foundations of quantum mechanics
% 03.65.Ge Solutions of wave equations: bound states
% 03.65.CA  Formalism
% 42.50.Dv Nonclassical ﬁeld states; squeezed, antibunched, and sub-Poissonian states

\maketitle %\maketitle must follow title, authors, abstract and \pacs

% Body of paper goes here. Use proper sectioning commands.
% References should be done using the \cite, \ref, and \label commands
% \section{Introduction}
%\label{}
% \subsection{}
% \subsubsection{}
\section{Introduction}
\label{intro}
\begin{figure}
	\centering
	\includegraphics[width=0.7\textwidth]{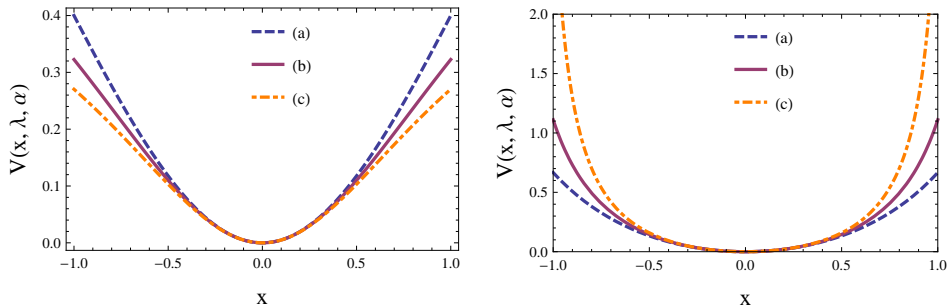}
	\caption{Plot of $V(x,\lambda,\alpha)$ as a function of $``x"$ and non-linearity parameter $``\lambda"$, for fixed value of $\alpha = 1$ with (a) $\lambda=0.25$, (b) $\lambda=0.55$, (c) $\lambda=0.85$ (left) and (a) $\lambda=-0.25$, (b) $\lambda=-0.55$, (c) $\lambda=-0.85$  (right).} \label{ppn}
\end{figure}

Many realistic phenomena in nature exhibit nonlinear oscillations which have motivated researchers to explore non-linear oscillators. Previously at classical level, Mathews and Lakshmanan \cite{ml75,lr03}, studied a non-linear differential equation
\begin{equation}\label{n3.1}
(1+\lambda x^{2})\ddot{x}-(\lambda x)\dot{x}^{2}+\alpha^{2}x=0,
\end{equation}
and it was shown that this equation admits periodic solutions of the form
\begin{equation}\label{n3.2}
x=A\sin(\omega t+\varphi),
\end{equation}
with $\omega=\frac{\alpha}{\sqrt{1+\lambda A^{2}}}$, which represent non-linear oscillations with quasi-harmonic form. Furthermore, it was shown that the dynamics of the non-linear harmonic oscillator, defined in Eq. (\ref{n3.1}), is governed by the Lagrangian
\begin{equation}\label{n3.3}
L=\frac{1}{2}\bigg[\frac{1}{1+\lambda x^{2}}\bigg](\dot{x}^{2}-\alpha^{2}x^{2}),
\end{equation}
which the authors considered as a one-dimensional analogue of a Lagrangian density appearing in some models of quantum field theory \cite{pcp,llv}. The Eq.~(\ref{n3.3}) represents the Lagrangian of oscillator with position-dependent effective mass (PDEM)
\begin{equation}\label{n3.4}
m(x)=\frac{1}{1+\lambda x^{2}}.
\end{equation}
From Eqs. (\ref{n3.3}) and (\ref{n3.4}), momentum of the non-linear oscillator can be written as
\begin{equation}\label{n3.5}
p=\frac{\partial L}{\partial\dot{x}}=\frac{\dot{x}}{1+\lambda x^{2}},
\end{equation}
which enables us to write the classical Hamiltonian as
\begin{equation}\label{ch}
H(x,\lambda,\alpha)=\frac{1}{2}(1+\lambda x^{2})p^{2}+V(x,\lambda,\alpha),
\end{equation}
where the potential term is of the following form
\begin{equation}\label{pe}
V(x,\lambda,\alpha)= \frac{1}{2}\bigg(\frac{\alpha^{2}x^{2}}{1+\lambda x^{2}}\bigg).
\end{equation}
The physical interpretation of this system is two-fold. On one hand, it represents the dynamics of a particle with spatially varying mass in one-dimensional space and on other hand, it can be considered as a Hamiltonian system describing the motion of a particle on the one-dimensional curved space defined by the metric $d\mu =\{m(x)\}^{\frac{1}{2}}dx$ and under the action of confining potential. Here it is important to note that, this $\lambda-$dependent system can be considered as a deformation to the linear harmonic oscillator in the sense that for $\lambda=0$ all the characteristics of the linear oscillator with constat mass are recovered. It is important to remark that this non-linearity parameter can take positive as well as negative values. Clearly, for negative values of $\lambda$, there exists a singularity for the mass function and associated dynamics at $1-|\lambda|x^{2} = 0$. Therefore, for $\lambda<0$, we shall restrict ourselves to the interior of the interval $x < 1/\sqrt{|\lambda|}$, where the kinetic energy term is positive definite. Fig. \ref{ppn} shows the $\lambda-$dependent potential energy for several values of $\lambda$.\\
\indent The theory of PDEM systems has attracted a lot of attentions due to the advent of sophisticated technologies to grow ultrathin semiconductor structures, with very prominent quantum effects. The quantum mechanical systems with spatially varying mass plays a vital role in the description of many systems \cite{llv,1,2,rm,3,4,4.1,4.2,5,7,7.1,7.2,7.3,7.4,o1,o3,o2,com,nse,nss,nsnew,ref1}. The qualitative understanding of a complicated realistic system can be acquired by analyzing the exact solutions of a related simplified model. Exact solutions to PDEM systems are of great importance due to the fact that they have numerous applications in the study of compositionally graded crystals \cite{1}, electronic properties of semiconductors \cite{2,rm}, quantum liquids \cite{3}, quantum dots \cite{4,4.1,4.2}, Helium clusters \cite{5}, semiconductor heterostructure \cite{2,7,7.1,7.2,7.3,7.4,o3} and the dependence of energy gap on magnetic field in semiconductor nano-scale quantum rings \cite{llv}. However, the quantization of position-dependent effective mass systems and finding their solutions, involve some conceptual and mathematical difficulties of a fundamental nature \cite{rm,o1,o2,com,nss,nsnew,ae70,befg}. For example, the quantization of a PDEM system experiences the ordering ambiguity of the operators concerning momentum and the spatially varying mass, involved in the kinetic energy term \cite{o1,o2,nse,nss,nsnew}.\\
\indent Inspired by the Schr\"{o}dinger factorization approach \cite{es40,ih}, the algebraic method \cite{es41,es4153,om,s1,s3,shdb,shdp} along with the concept of shape-invariance \cite{gsmt,b98,b99,s03,a04,s09,d88,c98,dutt88,l89,g,f,glo,hsa05,ab07,ab14} has become a power tool in finding solutions of exactly solvable quantum mechanical systems. The algebraic method provides us with a powerful tool for obtaining solutions and the underlying algebraic structure of the exactly solvable systems \cite{nsnew,b98,glo,hsa05,ab07}. The underlying algebra of a system has vast applications in different areas of mathematics and physics, such as it plays an important role in the theory of coherent states. Coherent states are extremely useful in various areas such as quantum mechanics, quantum optics, quantum information and group theory. \\
\indent Coherent states have attracted considerable attention in the literature \cite{a04,f,ab07,wg90,amp86,G.1,G.2,G.3,aagm,ajj,iqbal2010space,iqbal2011generalized,iqbal2012,iqbal2013gazeau}. The Glauber's formalism \cite{G.1,G.2,G.3} for the construction of the coherent states of the harmonic oscillator is based on Heisenberg-Weyl algebra. These states (CS) are defined as: (i) the eigenstates of the annihilation operator; (ii) the displaced vacuum states; (iii) the minimum uncertainty states. Due to numerous applications of the CS in mathematics and physics, the notion of CS have been generalized for the systems other than the harmonic oscillator \cite{a04,f,ab07,amp86,aagm,ajj}. In this regard Klauder \cite{klu.0} developed a generalized formalism to relate the quantum dynamics to the classical one. Initially, the concept of coherent states was generalized by using the algebraic structure of the pertaining system. In 1971, Barut and Girardello \cite{b.g} developed the CS for non-compact groups. These states were named as Barut-Girardello coherent states (BG CS). In the context of constant mass systems, a lot of contributions has been made by several authors \cite{a04,f,ab07,bd,dp,dp02,ha,hf,sm,dnvim,dsh,ba,hgs}. Particularly, different kinds of deformed nonlinear oscillators and their associated coherent states have been discussed extensively \cite{q1,f1}, as a deviation to linear harmonic oscillator. However, in the case of PDEM systems the generalization of CS has not yet been explored significantly except a few recent contributions \cite{nse,nss,m09,b09,r10,sco,sg12,sam}.\\
\indent In the present work we construct the BG CS of the non-linear harmonic oscillator with PDEM and study the basic properties of these CS. In Section 2, a self contained study of the ladder operators and associated algebraic structure of the non-linear harmonic oscillator with spatially varying mass is presented. Realization of dynamic group $SU(1,1)$ for the pertaining system is discussed in Section 3. Section 4, is dedicated to the construction of CS by using Barut-Girardello formalism and some basic properties such as continuity of parameters and resolution of unity are proved. In Section 5, some statistical properties such as Mandel parameter and the second order correlation \cite{grj07,mandel1,mandel2} are discussed which reflect the non-classical nature of these states. We close our work by some concluding remarks in Section 6.
%%%%%%%%%%%%%%%%%%%%%%%%%%%%%%%%%%%%%%%%%%%%%%%%%%%%%%%%%%%%%%%%%%%%%%%%%%%%%%%%%%%%%%%%%%%%%%%%%%%%
\section{Ladder operators and algebraic structure}\label{LO}
The classical Hamiltonian of the non-linear harmonic oscillator, given in Eq. (\ref{ch}), can be quantized by considering the symmetric ordering of the operators concerning momentum $p$ and spatially varying mass $m(x)$ \cite{o1,o2,nse,nss,nsnew,shdp} which provides us with a Hermitian Hamiltonian of the form
\begin{equation}\label{h'}
\hat{H}(\zeta,\tilde{\lambda},\alpha)=\frac{\alpha}{2}\bigg[-(1+\tilde{\lambda} \zeta^{2})\frac{d^{2}}{d\zeta^{2}}-2\tilde{\lambda} \zeta\frac{d}{d\zeta}+\frac{\zeta^{2}}{1+\tilde{\lambda} \zeta^{2}}\bigg],
\end{equation}
where we have used the dimensionless variables $\zeta=\sqrt{\alpha}x$ and $\tilde{\lambda}=\lambda/\alpha$. Following the general procedure of constructing the ladder operators for PDEM systems \cite{nsnew}, we introduce a pair of intertwining operators as
\begin{eqnarray}\label{3.2}
% \nonumber to remove numbering (before each equation)
\hat{A}(\zeta,\tilde{\lambda},\alpha) &=& \frac{1}{\sqrt{2}}\bigg[\sqrt{1+\tilde{\lambda} \zeta^{2}}\frac{d}{d\zeta}+\frac{\zeta}{\sqrt{1+\tilde{\lambda} \zeta^{2}}}\bigg], \nonumber \\
\hat{A}^{\dag}(\zeta,\tilde{\lambda},\alpha) &=& \frac{1}{\sqrt{2}}\bigg[-\sqrt{1+\tilde{\lambda} \zeta^{2}}\frac{d}{d\zeta}+\frac{(1 - \tilde{\lambda}) \zeta}{\sqrt{1+\tilde{\lambda} \zeta^{2}}}\bigg].
\end{eqnarray}
Note that, in order to avoid notational complexity, we will suppress the $\zeta-$dependence and $\tilde{\lambda}-$dependance of all the operators in the ongoing analysis, for instance, we will use $\hat{A}(\alpha)$ instead of $\hat{A}(\zeta,\tilde{\lambda},\alpha)$. The intertwining operators defined in Eq. (\ref{3.2}), are constructed in such a way that 
\begin{equation}\label{gsc}
\hat{A}(\alpha)|\varphi_{0}\rangle=0.
\end{equation}
%and the ground state wave function of the Hamiltonian (\ref{h}), takes the form
%\begin{equation}\label{3.1}
%\varphi_{0}(\zeta)=(1+\tilde{\lambda} \zeta^{2})^{\frac{-1}{2\tilde{\lambda}}}.
%\end{equation}
Let us consider the product of the operators $\hat{A}(\alpha)$ and $\hat{A}^{\dag}(\alpha)$ as
\begin{eqnarray}\label{3.3}
\hat{A}^{\dag}(\alpha)\hat{A}(\alpha)=\frac{1}{2}\left[
\begin{array}{c}
-(1+\tilde{\lambda} \zeta^{2})\frac{d^{2}}{d\zeta^{2}}-2\tilde{\lambda} \zeta\frac{d}{d\zeta}
+\frac{\zeta^{2}}{1+\tilde{\lambda} \zeta^{2}}  \\
\end{array}
\right]-\frac{1}{2},
\end{eqnarray}
which leads to 
\begin{equation}\label{hc}
\hat{H}=\alpha\bigg(\hat{A}^{\dag}(\alpha)\hat{A}(\alpha)+\frac{1}{2}\bigg).
\end{equation}
Thus, the operators introduced in Eq. (\ref{3.2}), factorizes the Hamiltonian (\ref{h'}) and provides us with the ground states energy of the given systems as $ E_{0}=\frac{\alpha}{2}.$ We now consider the product of intertwining operators introduced in Eq. (\ref{3.2}), as
\begin{equation}\label{3.4}
\hat{A}(\alpha)\hat{A}^{\dag}(\alpha)=\frac{1}{2}\left[
\begin{array}{c}
-(1+\tilde{\lambda} \zeta^{2})\frac{d^{2}}{d\zeta^{2}}-2\tilde{\lambda} \zeta\frac{d}{d\zeta}
+\frac{(1-\tilde{\lambda})^{2}\zeta^{2}}{1+\tilde{\lambda} \zeta^{2}}
\end{array}
\right]+\frac{1}{2}(1-\tilde{\lambda}).
\end{equation}
In order to specify the shape-invariance condition \cite{nsnew,b98,a04,f,ab07,ab14}
\begin{equation}\label{3.6'}
\hat{A}(\alpha_{1})\hat{A}^{\dag}(\alpha_{1})=\hat{A}^{\dag}(\alpha_{2})\hat{A}(\alpha_{2})+R(\alpha_{1}),
\end{equation}
we make use of Eqs. (\ref{3.3}) and (\ref{3.4}) in Eq. (\ref{3.6'}) and get
\begin{eqnarray}\label{3.6}
R(\alpha_{1})=\hat{A}(\alpha_{1})\hat{A}^{\dag}(\alpha_{1})-\hat{A}^{\dag}(\alpha_{2})\hat{A}(\alpha_{2})=\alpha_{1}-\tilde{\lambda},
\end{eqnarray}
where the parameters concerning shape-invariance are related as
\begin{eqnarray}\label{3.7}
\alpha_{1}&=&1, \nonumber \\
\alpha_{n}&=&\alpha_{n-1}-\tilde{\lambda}=1-(n-1)\tilde{\lambda}, \nonumber \\
R(\alpha_{n})&=&R(\alpha_{n-1})-\tilde{\lambda}=1-n\tilde{\lambda}.
\end{eqnarray}
It is important to note that for the present case, the parameters $\alpha_{1}$ and $\alpha_{2}$ are related by means of a translation $\alpha_{2}=\alpha_{1}-\tilde{\lambda}$, where $``-\tilde{\lambda}"$ is the translation step \cite{nsnew}. This reparametrization of the set of parameters $\alpha_{1}$, with the set $\alpha_{2}$, is achieved by means of a similarity transformation
\begin{equation}\label{st}
\hat{T}(\alpha_{1})\hat{O}(\alpha_{1})\hat{T}^{-1}(\alpha_{1})=\hat{O}(\alpha_{2}),
\end{equation}
where $\hat{T}(\alpha_{1})$ is a translation operator \cite{nsnew,b98,f,hsa05,ab14} defined as
\begin{equation}\label{3.8}
% \nonumber to remove numbering (before each equation)
\hat{T}(\alpha_{1})|\varphi(\alpha_{1})\rangle=|\varphi(\alpha_{2})\rangle.
\end{equation}
In order to establish the algebraic structure of the quantum mechanical system with PDEM \cite{nsnew}, we introduce appropriate ladder operators by using the intertwining operators given in Eq. (\ref{3.2}) and the translation operators in Eq. (\ref{3.8}), as
\begin{equation}\label{3.9}
% \nonumber to remove numbering (before each equation)
\hat{L}_{-} = \hat{T}^{-1}\hat{A}~~\mbox{and}~~\hat{L}_{+}=\hat{A}^{\dag}\hat{T},
\end{equation}
whose action as the annihilation operator and the creation operator is defined respectively as \cite{nsnew}
\begin{eqnarray}\label{3.10}
% \nonumber to remove numbering (before each equation)
\hat{L}_{-}|\varphi_{n}\rangle &=& \sqrt{n-\frac{\tilde{\lambda}}{2}n(n+1)}
~~|\varphi_{n-1}\rangle, \\
\nonumber \hat{L}_{+}|\varphi_{n}\rangle &=& \sqrt{(n+1)-\frac{\tilde{\lambda}}{2}(n+1)(n+2)}
~~|\varphi_{n+1}\rangle.
\end{eqnarray}
Note that in terms of these ladder operators, Eq. (\ref{gsc}), takes the form $\hat{L}_{-}|\varphi_{0}\rangle=0,$
%\begin{equation}\label{gsc1}
%\hat{L}_{-}|\varphi_{0}\rangle=0,
%\end{equation}
and (\ref{hc}) becomes
\begin{equation}\label{hc1}
\hat{H}=\alpha\bigg(\hat{H}_{1}+\frac{1}{2}\bigg),
\end{equation}
where $\hat{H}_{1}=\hat{L}_{+}\hat{L}_{-}$. With the help of these ladder operators, the underlying algebra turns to be
\begin{eqnarray}\label{3.13}
\nonumber  [\hat{L}_{-},~\hat{L}_{+}]=R(\alpha_{0})=1,~~
\nonumber [\hat{L}_{\mp},~R(\alpha_{0})]=\pm \tilde{\lambda}~\hat{L}_{\mp},\\
\end{eqnarray}
Here, the underlying algebra for the shape-invariant potential is finite-dimensional. The eigenstates of the non-linear PDEM oscillator are given as
%\begin{equation}\label{3.11}
% E_{n} = \alpha \bigg[ n+\frac{1}{2} - \bigg(\frac{\tilde{\lambda}}{2}\bigg)n(n+1)\bigg],
%\end{equation}
%and
\begin{eqnarray}\label{3.11'}
|\varphi_{n}\rangle= \frac{[\hat{L}_{+}]^{n}|\varphi_{0}\rangle}
{\sqrt{[n]!}},
\end{eqnarray}
where $[n]!$ is the generalized factorial which can be written in terms of $R_{\alpha_{n}}$, defined in Eq. (\ref{3.7}), as
\begin{eqnarray}\label{facho}
[n]!&=&[R(\alpha_{n})+R(\alpha_{n-1})+...+R(\alpha_{1})][R(\alpha_{n-1})+...+R(\alpha_{1})]...[R(\alpha_{1})],\nonumber\\
&=&(-1)^{n}~n!\bigg(\frac{\tilde{\lambda}}{2}\bigg)\bigg(2-\frac{2}{\tilde{\lambda}}\bigg)_{n}.
%&=&\bigg(\frac{-\tilde{\lambda}}{2}\bigg)^{n}
%\frac{n!\Gamma\big(2-\frac{2}{\tilde{\lambda}}+n\big)}{\Gamma\big(2-\frac{2}{\tilde{\lambda}}\big)}.
\end{eqnarray}
Here $(x)_{n}=x(x+1)\dots(x+n-1)$ represents the Pochhammer symbol.
%%%%%%%%%%%%%%%%%%%%%%%%%%%%%%%%%%%%%%%%%%%%%%%%%%%%%%%%%%%%%%%%%%%%%%%%%%%%%%%%%%%%%%%%%%%%%%%%%%%%%%%%
%%%%%%%%%%%%%%%%%%%%%%%%%%%%%%%%%%%%%%%%%%%%%%%%%%%%%%%%%%%%%%%%%%%%%%%%%%%%%%%%%%%%%%%%%%%%%%%%%%%%%%%%
\section{Realization of dynamic group $SU(1,1)$}\label{RDGSU}
As shown in the previous section that the non-linear harmonic oscillator with position-dependent effective mass posses a finite-dimensional Lie algebra. This algebraic structure not only enables us to determine the spectrum of the system but it also provides us with the dynamic group of the underlying system. Since the Lie group obtained in this way has a involved algebraic structure, thus based on Barut's formalism \cite{shdb,barut}, we construct the following operators
\begin{eqnarray}\label{no}
% \nonumber to remove numbering (before each equation)
\hat{K}_{-} &=& \hat{L}_{-}\sqrt{\hat{H}_{1}}, \nonumber \\
\hat{K}_{+} &=& \sqrt{\hat{H}_{1}}~\hat{L}_{+}, \nonumber \\
\hat{K}_{0}&=& \frac{1}{2}+\hat{H}_{1},
\end{eqnarray}
where $\hat{H}_{1}=\hat{L}_{+}\hat{L}_{-}$. The newly constructed operators given in Eq. (\ref{no}), satisfy the following commutation relations
\begin{eqnarray}\label{3.13}
\nonumber  [\hat{K}_{-},~\hat{K}_{+}]=2\hat{K}_{0},~~
{[\hat{K}_{0},~\hat{K}_{\pm}]}=\pm \hat{K}_{\pm},
\end{eqnarray}
which corresponds to the commutation relations of the generators of Lie algebra $su(1,1)$. Thus, the dynamic group of the non-linear harmonic oscillator with position-dependent effective mass is the non-compact group $SU(1,1)$. \\
\indent The Casimir operator in this case turns out to be
\begin{eqnarray}\label{cas}
\nonumber  C=\hat{K}_{+}\hat{K}_{-}-\hat{K}_{0}(\hat{K}_{0}-1)
= \frac{1}{4}.
\end{eqnarray}
In terms of the newly constructed lowering operator defined in Eq. (\ref{no}), we get
\begin{equation}\label{gsc2}
\hat{K}_{-}|\varphi_{0}\rangle=0.
\end{equation}
Also, Eqs.(\ref{no}) together with (\ref{3.10}), yield
\begin{eqnarray}\label{3.10'}
% \nonumber to remove numbering (before each equation)
\hat{K}_{0}|\varphi_{n}\rangle &=& \bigg[n+\frac{1}{2}-\lambda^{'}n(n+1)\bigg]\varphi_{n}\rangle,\\
\nonumber\hat{K}_{-}|\varphi_{n}\rangle &=& [n-\lambda^{'}n(n+1)]\varphi_{n-1}\rangle, \\
\nonumber \hat{K}_{+}|\varphi_{n}\rangle &=& [(n+1)-\lambda^{'}(n+1)(n+2)]|\varphi_{n+1}\rangle,
\end{eqnarray}
where $\lambda^{'}=\frac{\tilde{\lambda}}{2}$.\\
\indent It is well known that there are four irreducible representations for the $su(1,1)$ Lie algebra \cite{shdb,gbm}. It is clear from Eq. (\ref{gsc2}) that there exists a ground state for the non-linear harmonic oscillator with position-dependent effective mass. Thus for the present case, among these four representations if the dynamic group $SU(1,1)$, we choose the irreducible representation $D^{+}(j)$ with respect to the basis vectors $|m,j\rangle$,
\begin{eqnarray}\label{rc}
% \nonumber to remove numbering (before each equation)
M_{0}|m,j\rangle &=& m|m,j\rangle, \nonumber \\
M_{-}|m,j\rangle &=& \sqrt{(m+j)(m-j-1)}|m-1,j\rangle, \nonumber \\
M_{+}|m,j\rangle &=& \sqrt{(m+j+1)(m-j)}|m+1,j\rangle, \\
m&=&k-j,~~~ j<0,~~~~ k=0,1,2,...., \nonumber
\end{eqnarray}
for which the energy spectrum is bounded from below. Comparison of Eq. (\ref{3.10'}) with Eq. (\ref{rc}), provides us with $m=n+\frac{1}{2}-\lambda^{'}n(n+1)$ and $j=\frac{-1}{2}$, so that
$|\varphi_{n}\rangle=|n+\frac{1}{2}-\lambda^{'}n(n+1),\frac{-1}{2}\rangle$.
%%%%%%%%%%%%%%%%%%%%%%%%%%%%%%%%%%%%%%%%%%%%%%%%%%%%%%%%%%%%%%%%%%%%%%%%%%%%%%%%%%%%%%%%%%%%%%%%
%%%%%%%%%%%%%%%%%%%%%%%%%%%%%%%%%%%%%%%%%%%%%%%%%%%%%%%%%%%%%%%%%%%%%%%%%%%%%%%%%%%%%%%%%%%%%%%%
\section{Barut-Girardello coherent states}
Since the dynamic group of the non-linear harmonic oscillator with position-dependent effective mass is the Lie group $SU(1,1)$, thus, in this sense we can define the Barut-Girardello coherent state (BG CS) for the underlying system as
\begin{equation}\label{bcs}
K_{-}|z\rangle=z|z\rangle,
\end{equation}
where $K_{-}$ is the lowering operator defined in (\ref{no}). These states can be written as a linear combination of the eigenstates (\ref{3.11'}), of the Hamiltonian $\hat{H}$, as
\begin{equation}\label{lc}
|z\rangle=\sum_{n=0}^{\infty} c_{n}|\varphi_{n}\rangle.
\end{equation}
Using (\ref{3.10'}) and (\ref{bcs}) along with (\ref{lc}), we get
\begin{equation}
z \sum_{n=0}^{\infty} c_{n}|\varphi_{n}\rangle=\sum_{n=0}^{\infty} c_{n}[n-\lambda^{'}n(n+1)]|\varphi_{n-1}\rangle,
\end{equation}
which leads to the following recursion relation
\begin{equation}
c_{n}=\frac{z}{n[1-\lambda^{'}(n+1)]}c_{n-1}.
\end{equation}
Using above we get
\begin{equation}
c_{n}=\frac{z^{n}}{n!~(1-2\lambda^{'})(1-3\lambda^{'})...(1-(n+1)\lambda^{'})}c_{0},
\end{equation}
which can be rewritten in terms of gamma function as
\begin{equation}\label{rr}
c_{n}=(-1)^{n}~\frac{\Gamma(2-\frac{1}{\lambda^{'}})}{n!~\Gamma(2-\frac{1}{\lambda^{'}}+n)}
\bigg(\frac{z}{\lambda^{'}}\bigg)^{n}c_{0}.
\end{equation}
Using (\ref{rr}) in (\ref{lc}), we get
\begin{equation}\label{bgcs}
|z\rangle=\frac{1}{\sqrt{N(|z|^{2})}}\sum_{n=0}^{\infty} (-1)^{n}~\frac{\Gamma(2-\frac{1}{\lambda^{'}})}{n!~\Gamma(2-\frac{1}{\lambda^{'}}+n)}
\bigg(\frac{z}{\lambda^{'}}\bigg)^{n}|\varphi_{n}\rangle,
\end{equation}
where the normalization factor $c_{0}=\frac{1}{\sqrt{N(|z|^{2})}}$ can be calculated by using the normalization condition $\langle z|z \rangle=1$, as
\begin{equation}\label{nco}
N(|z|^{2})=~_{0}F_{3}\bigg(1,2-\frac{1}{\lambda^{'}},2-\frac{1}{\lambda^{'}}
;\frac{|z|^{2}}{(\lambda^{'})^{2}}\bigg).
\end{equation}
Here $_{0}F_{3}(a,b,c;\zeta)$ represents the hypergeometric function. \\
\indent Note that the BG CS defined in Eq. (\ref{bgcs}), satisfies the Klauder's minimal set of conditions \cite{klu.0}, that are required for any state to be coherent state. The overlap of two BG CS for the non-linear oscillator is given as
\begin{equation}\label{ol}
\langle z|z^{'} \rangle=\frac{1}{\sqrt{ N(|z|^{2}) N(|z^{'}|^{2})}}
~_{0}F_{3}\bigg(1,2-\frac{1}{\lambda^{'}},2-\frac{1}{\lambda^{'}}
;\frac{z^{'} z^{*}}{(\lambda^{'})^{2}} \bigg),
\end{equation}
from which it follows that the BG CS for the non-linear oscillator with position-dependent effective mass are not orthogonal. The continuity in the label $z$ follows immediately due to the fact that
\begin{equation}\label{cl}
\lim_{z^{'}\rightarrow z} \parallel |z^{'}\rangle-|z\rangle\parallel^{2}=
\lim_{z^{'}\rightarrow z} [2(1-Re \langle z^{'}|z\rangle)]=0.
\end{equation}
We now investigate the over-completeness of the BG CS. For this we assume that there exist a positive and unique weight function $w(|z|^{2})$, such that
\begin{equation}\label{oc1}
\int d^{2}z w(|z|^{2}) |z\rangle\langle z|=1=\sum_{n=0}^{\infty}|\varphi_{n}\rangle\langle\varphi_{n}|.
\end{equation}
Substituting (\ref{bgcs}) in (\ref{oc1}) and introducing the variables $z=re^{i\theta}$ and $|z|^{2}=\xi$, we finally arrive at
\begin{equation}\label{oc2}
\int_{0}^{\infty}  \tilde{w}(\xi)\xi^{n} d\xi = \frac{\Gamma(n+1)\Gamma(n+1)\Gamma(2-\frac{1}{\lambda^{'}}+n)\Gamma(2-\frac{1}{\lambda^{'}}+n)}
{\Gamma(2-\frac{1}{\lambda^{'}})\Gamma(2-\frac{1}{\lambda^{'}})}(\lambda^{'})^{2n},
\end{equation}
where $\tilde{w}(\xi)=\frac{\pi w(\xi)}{N(\xi)}$. Our aim is to obtain the weight function which can be determined by using inverse Mellin transform. It is well known that the Mellin transform of the Meijer's $G$-function \cite{rkam}, is given as
\begin{eqnarray}\label{oc3}
\nonumber \int_{0}^{\infty} G_{p,q}^{m,n}\bigg(  \begin{array}{c}
a_{1},...a_{n},a_{n+1},...a_{p} \\
b_{1},...b_{m},b_{m+1},...b_{q}
\end{array}\bigg|\beta \xi
\bigg) \xi^{k-1} d\xi=\\
\frac{\prod_{i=1}^{m}\Gamma(b_{i}+k)\prod_{i=1}^{n}\Gamma(1-a_{i}-k)}
{\prod_{i=m+1}^{q}\Gamma(1-b_{i}-k)\prod_{i=n+1}^{p}\Gamma(a_{i}+k)}(\beta)^{-k}.
\end{eqnarray}
Comparison of (\ref{oc2}) and (\ref{oc3}), provides us with the required weight function $w(\xi)$, as
\begin{equation}\label{oc4}
w(\xi)=\frac{~_{0}F_{3}\bigg(1,2-\frac{1}{\lambda^{'}},2-\frac{1}{\lambda^{'}}
	;\frac{\xi}{(\lambda^{'})^{2}}\bigg)}
{\pi[\lambda^{'}~\Gamma(2-\frac{1}{\lambda^{'}})]^{2}}
G_{0,4}^{4,0}\bigg(  \begin{array}{c}
.... \\
0,0,1-\frac{1}{\lambda^{'}},1-\frac{1}{\lambda^{'}}
\end{array}\bigg| \frac{\xi}{(\lambda^{'})^{2}}
\bigg),
\end{equation}
which satisfies the integral equation (\ref{oc1}). The radius of convergence for the non-linear harmonic oscillator with position-dependent effective mass is given as
\begin{equation}\label{oc5}
R=\lim_{n\rightarrow \infty} \bigg[(-\lambda^{'})^{n}~\frac{n!~\Gamma(2-\frac{1}{\lambda^{'}}+n)}{\Gamma(2-\frac{1}{\lambda^{'}})}\bigg]^{\frac{1}{n}}
=\infty.
\end{equation}
This shows that the BG CS are defined on the whole complex plane.
%%%%%%%%%%%%%%%%%%%%%%%%%%%%%%%%%%%%%%%%%%%%%%%%%%%%%%%%%%%%%%%%%%%%%%%%%%%%%%%%%%%%%%%%%%%%%%%%
%%%%%%%%%%%%%%%%%%%%%%%%%%%%%%%%%%%%%%%%%%%%%%%%%%%%%%%%%%%%%%%%%%%%%%%%%%%%%%%%%%%%%%%%%%%%%%%%
\section{Statistical properties}\label{SP}
The statistical features of a coherent state can be characterized by
inherent probability to occupation the $nth$ eigenstate in the coherent superposition constituting the coherent state.
For the BG CS, introduced in (\ref{bgcs}),the probability distribution is given by
{\begin{equation}\label{pd}
	% \nonumber to remove numbering (before each equation)
	P_{n}=\vert c_{n}\vert^{2} = \frac{1}{ N(|z|^{2})}
	\bigg[\frac{\Gamma(2-\frac{1}{\lambda^{'}})}{n!~\Gamma(2-\frac{1}{\lambda^{'}}+n)}\bigg]^{2}
	\bigg(\frac{|z|}{\lambda^{'}}\bigg)^{2n},
	\end{equation}
	which is plotted, in Fig. (\ref{distribution}).
	\begin{figure}
		\centering
		\includegraphics[width=0.94\textwidth]{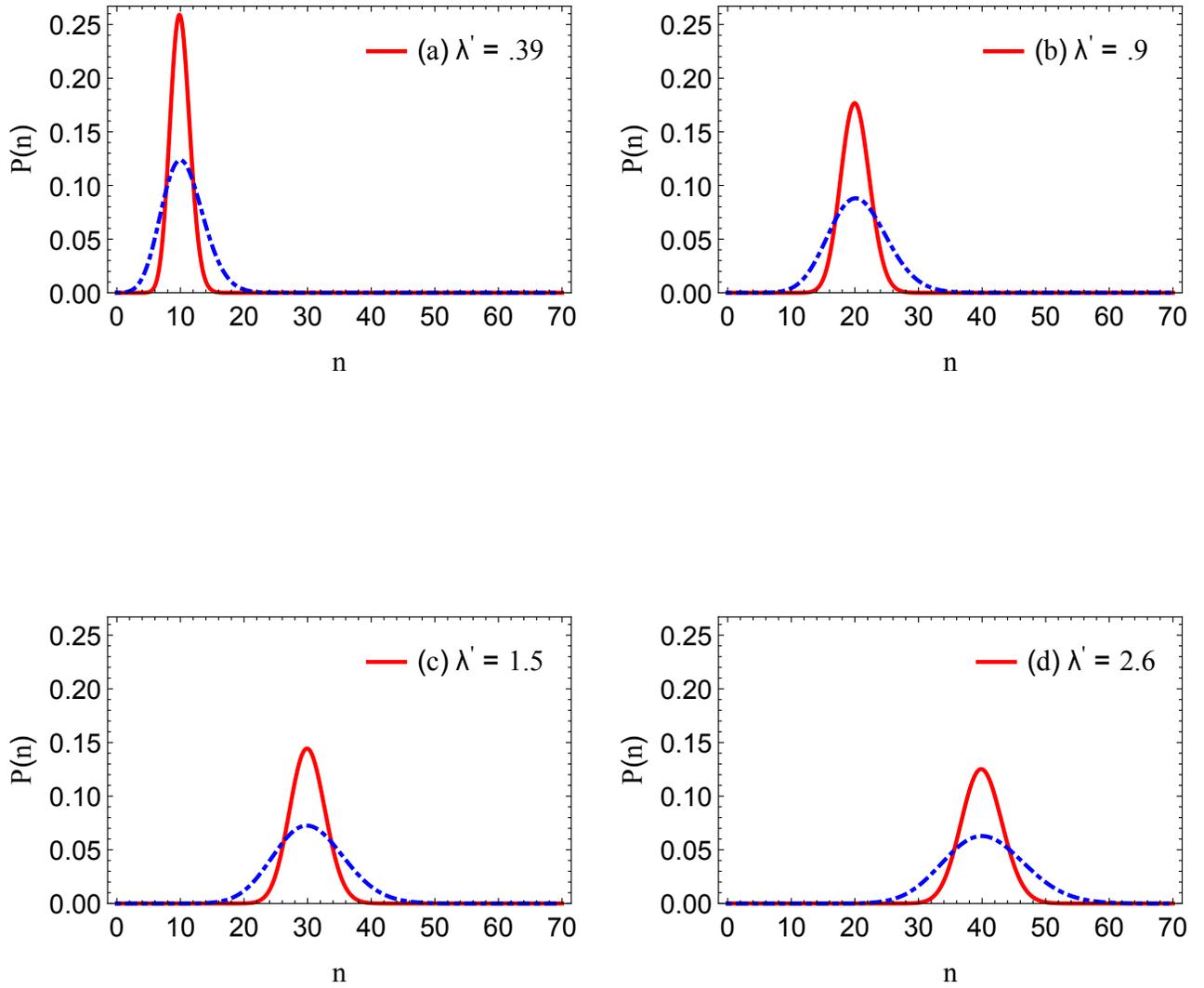}
		\caption{The weighting distribution $P(n)$ for the non-linear harmonic oscillator (solid curve) and the linear harmonic oscillator (dashed curve) as a function of quantum number $n$ for (a) $\lambda^{'}=0.39$, (b) $\lambda^{'}=.9$, (c) $\lambda^{'}=1.5$ and (d) $\lambda^{'}=2.6$.} \label{distribution}
	\end{figure}
	It is clear from the figure that the this distribution is narrower than the weighting distribution for the harmonic oscillator. This indicates the sub-Poissonian nature of the distribution for the non-linear oscillator with spatially varying mass.\\
	It is a well known fact that a probability distribution is characterized by its mean and the corresponding variance, which can be calculated by using the first and second moments of the probability. For the distribution (\ref{pd}), the first moment is given as
	\begin{eqnarray}\label{fm}
	% \nonumber to remove numbering (before each equation)
	\langle n\rangle= \sum_{n=0}^{\infty} nP_{n}
	= \frac{1}{ N(|z|^{2})}\bigg(\frac{|z|}{2\lambda^{'}-1}\bigg)^{2}
	~_{0}F_{3}\bigg(2,3-\frac{1}{\lambda^{'}},3-\frac{1}{\lambda^{'}};\frac{|z|^{2}}{(\lambda^{'})^{2}}\bigg),
	\end{eqnarray}
	which provides us with the mean of the distribution. The second moment is calculated as,
	\begin{eqnarray}\label{sm}
	% \nonumber to remove numbering (before each equation)
	\langle n^{2}\rangle = \sum_{n=0}^{\infty} n^{2}P_{n}
=\frac{1}{ N(|z|^{2})}\bigg(\frac{|z|}{2\lambda^{'}-1}\bigg)^{2}
	~_{0}F_{3}\bigg(1,3-\frac{1}{\lambda^{'}},3-\frac{1}{\lambda^{'}};\frac{|z|^{2}}{(\lambda^{'})^{2}}\bigg).
	\end{eqnarray}
	Hence, the Eqs. (\ref{fm}) and (\ref{sm}) lead us to calculate the variance of the probability distribution as
	\begin{eqnarray}\label{v}
	(\Delta n)^{2} &=& \langle n^{2}\rangle-(\langle n\rangle) ^{2}, \nonumber\\
	&=&  \frac{|z|^{2}}{(2\lambda^{'}-1)^{2}~ N(|z|^{2})}
	\left[
	\begin{array}{c}
	~_{0}F_{3}\bigg(1,3-\frac{1}{\lambda^{'}},3-\frac{1}{\lambda^{'}};\frac{|z|^{2}}{(\lambda^{'})^{2}}\bigg)- \\
	\frac{|z|^{2}}{(2\lambda^{'}-1)^{2}~ N(|z|^{2})}
	~_{0}F^{2}_{3}\bigg(2,3-\frac{1}{\lambda^{'}},3-\frac{1}{\lambda^{'}};\frac{|z|^{2}}{(\lambda^{'})^{2}}\bigg)
	\\
	\end{array}
	\right].
	\end{eqnarray}
	\begin{figure}
		\centering
		\includegraphics[width=0.7\textwidth]{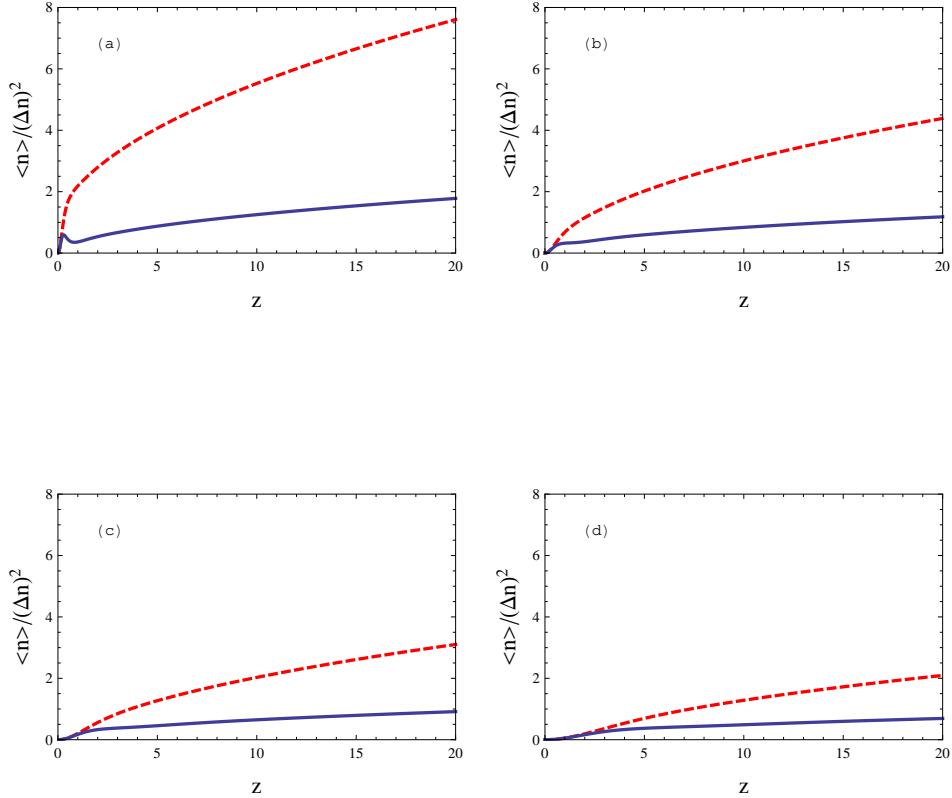}
		\caption{The mean $\langle n\rangle$ (dashed curve) and the variance $(\Delta n)^{2}$ (solid curve) as a function of the coherent state parameter $z$ for (a) $\lambda^{'}=0.39$, (b) $\lambda^{'}=.9$, (c) $\lambda^{'}=1.5$ and (d) $\lambda^{'}=2.6$.} \label{mvbg}
	\end{figure}
	In Fig. (\ref{mvbg}), we show the graph of mean and the variance for the non-linear harmonic oscillator as a function of the coherent state parameter ``$z$". This clearly indicates that the distribution is not Poissonian in the present case.\\
	\indent In general, the Mandel parameter is an indispensable tool to determine the nature of a weighting distribution. It is defined \cite{grj07,mandel1,mandel2} as
$Q=\frac{(\Delta n)^{2}-\langle n\rangle}{\langle n\rangle},$ where the weighting distribution of a coherent state is Poissonian if $Q = 0$, super-Poissonian if $Q > 0$ and sub-Poissonian if $Q < 0$. Substitution of Eqs. (\ref{fm}) and (\ref{v}), in Eq. (\ref{mp}), yield
	\begin{equation}\label{mp}
	Q=\frac{~_{0}F_{3}\bigg(1,3-\frac{1}{\lambda^{'}},3-\frac{1}{\lambda^{'}};\frac{|z|^{2}}{(\lambda^{'})^{2}}\bigg)}
	{~_{0}F_{3}\bigg(2,3-\frac{1}{\lambda^{'}},3-\frac{1}{\lambda^{'}};\frac{|z|^{2}}{(\lambda^{'})^{2}}\bigg)}-
	\frac{|z|^{2}/(2\lambda^{'}-1)^{2}}{ N(|z|^{2})}
	~_{0}F_{3}\bigg(2,3-\frac{1}{\lambda^{'}},3-\frac{1}{\lambda^{'}};\frac{|z|^{2}}{(\lambda^{'})^{2}}\bigg)-1.
	\end{equation}
	Another important parameter that provides information on the bunching or the antibunching effects is the second-order correlation function \cite{grj07,mandel1,mandel2}, which is defined as
$g^{2}(0)=\frac{\langle n^{2}\rangle-\langle n\rangle}{(\langle n\rangle)^{2}}=\frac{Q}{\langle n\rangle}+1.$
	If $g^{2}(0)<1 ~(g^{2}(0)>1)$, the anti-bunching (bunching) effect appears. The case $g^{2}(0)=1$ corresponds to the coherent states of the harmonic oscillator. Now for the system under consideration the second-order correlation function turns out to be
	\begin{equation}\label{socb2}
	g^{2}(0)=\frac{(2\lambda^{'}-1)^{2}N(|z|^{2})}{|z|^{2}
		~_{0}F_{3}\bigg(2,3-\frac{1}{\lambda^{'}},3-\frac{1}{\lambda^{'}};\frac{|z|^{2}}{(\lambda^{'})^{2}}\bigg)}
	\bigg[
	\frac{~_{0}F_{3}\bigg(1,3-\frac{1}{\lambda^{'}},3-\frac{1}{\lambda^{'}};\frac{|z|^{2}}{(\lambda^{'})^{2}}\bigg)}
	{~_{0}F_{3}\bigg(2,3-\frac{1}{\lambda^{'}},3-\frac{1}{\lambda^{'}};\frac{|z|^{2}}{(\lambda^{'})^{2}}\bigg)}-1
	\bigg],
	\end{equation}
	which indicates that BG CS of the non-linear harmonic oscillator with position-dependent effective mass, exhibit the anti-bunching effect. Fig. \ref{mqp}(a) represents the plot of Mandel parameter as a function of the coherent state parameter $``z"$ and it clearly indicates the sub-Poissonian nature of the distribution for the BG CS of the confining system. The graph of second-order correlation function in Fig. \ref{mqp}(b) clearly indicates the anti-bunching phenomenon.
	\begin{figure}
		\centering
		\includegraphics[width=.75\textwidth]{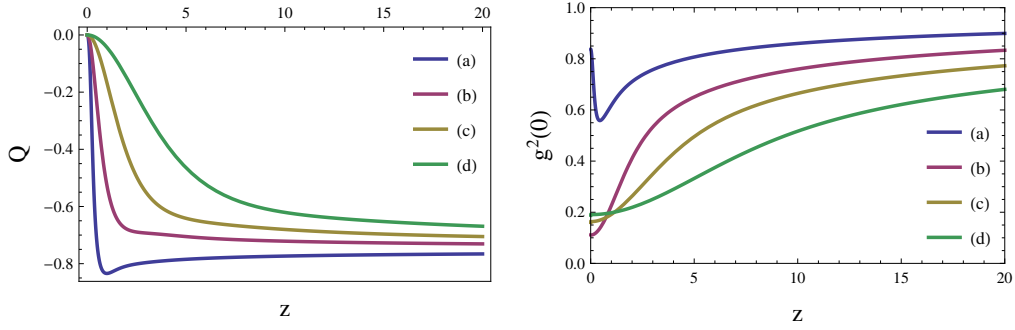}
		\caption{Mandel parameter $Q$ (left) and the second-order correlation function (right) as a function of the coherent state parameter $z$ for (a) $\lambda^{'}=0.39$, (b) $\lambda^{'}=.9$, (c) $\lambda^{'}=1.5$ and (d) $\lambda^{'}=2.6$.} \label{mqp}
	\end{figure}
	%%%%%%%%%%%%%%%%%%%%%%%%%%%%%%%%%%%%%%%%%%%%%%%%%%%%%%%%%%%%%%%%%%%%%%%%%%%%%%%%%%%%%%%%%%%%%%%%%%%%%%%%%%%%
	%%%%%%%%%%%%%%%%%%%%%%%%%%%%%%%%%%%%%%%%%%%%%%%%%%%%%%%%%%%%%%%%%%%%%%%%%%%%%%%%%%%%%%%%%%%%%%%%%%%%%%%%%%%%
	\section{Conclusion}\label{C}
	In the context of the generalized CS, a non-linear harmonic oscillator with PDEM has been studied. The ladder operators along with the associated algebra have been explicitly obtained which provide a strong basis for the construction of algebraic dependent CS. In order to identify the dynamic group of the system under consideration, we introduced a set of operators which satisfies the algebraic structure of the non-compact group $SU(1,1)$. This enables us to construct the generalized coherent states using Barut Girardello formalism. It has been shown that these states satisfy the basic set of conditions required for any state to be a coherent state. \\
	\indent Various statistical properties of these states have also been discussed. The closed form relations for the Mandel parameter and second order correlation function are obtained and it is shown that the BG CS possess a sub-Poissonian statistics and exhibits anti-bunching effect. Moreover, the harmonic limit for our analysis of the non-linear oscillator is obtained for $\lambda = 0$, such that all the results obtained for the non-linear harmonic oscillator with PDEM reduce to the corresponding results of the harmonic oscillator with constant mass.
	%%%%%%%%%%%%%%%%%%%%%%%%%%%%%%%%%%%%%%%%%%%%%%%%%%%%%%%%%%%%%%%%%%%%%%%%%%%%%
%%%%%%%%%%%%%%%%%%%%%%%%%%%%%%%%%%%%%%%%%%%%%%%%%%%%%%%%%%%%%%%%%%%%%%%%%%%%%%%%%%%%%%%%%%%%%%%%%%%%
%%%%%%%%%%%%%%%%%%%%%%%%%%%%%%%%%%%%%%%%%%%%%%%%%%%%%%%%%%%%%%%%%%%%%%%%%%%%%
%\section*{References}

%\bibliography{mybib}

\end{document}